\documentclass
[preprint,pre,nofootinbib,showpacs,titlepage,twocolumn,tightenlines,10pt]{revtex4}%
\usepackage{amsfonts}
\usepackage{amsmath}
\usepackage{amssymb}
\usepackage{graphicx}
\usepackage{acronym}%
\providecommand{\U}[1]{\protect\rule{.1in}{.1in}}
\providecommand{\U}[1]{\protect\rule{.1in}{.1in}}
\providecommand{\U}[1]{\protect\rule{.1in}{.1in}}

\begin{document}
\title{Self-Sustaining Oscillations in Complex Networks of Excitable Elements}
\author{Patrick McGraw and Michael Menzinger}
\affiliation{Department of Chemistry, University of Toronto, Toronto, Ontario, Canada M5S 3H6.}
\date{\today}

\begin{abstract}
Random networks of symmetrically coupled, excitable elements can self-organize
into coherently oscillating states if the networks contain loops (indeed loops
are abundant in random networks) and if the initial conditions are
sufficiently random. In the oscillating state, signals propagate in a single
direction and one or a few network loops are selected as driving loops in
which the excitation circulates periodically. We analyze the mechanism,
describe the oscillating states, identify the pacemaker loops and explain key
features of their distribution. \ This mechanism may play a role in epileptic
seizures. \ 

\end{abstract}
\maketitle

The coherent oscillation (CO) of a collection of units that are
non-oscillatory on their own is relevant to biological and physical
sciences:\ CO has been identified and analyzed in populations of excitable
biological cells (yeast\cite{DeMonte-yeast}, $\beta$ pancreatic
cells\cite{Cartwright-pancreas2000}, Dictyostelium
discoideum\cite{Nanjiundiah-Dictyostelium} and cultured heart
cells\cite{Bub-heart cells}) and of excitable catalytic
particles\cite{Tinsley2010}\cite{Taylor2009}\cite{Showalter}.\ \ In contrast
with well-studied synchronization phenomena of self-oscillating
units\cite{Sync},  in these cases the ability to oscillate derives from the
interactions of the elements. CO can occur also on complex networks if the
nodes are excitable \cite{Liao}\cite{Lewis} or even monostable\cite{Liao},
\ provided that the network contains \emph{loops}, and that the directional
symmetry of couplings is somehow broken to allow a signal to propagate in a
single direction around a loop\cite{Lewis}. \ Networks of these types include
some neural \cite{Marder}\cite{Selverston} and genetic regulatory
networks.\cite{Liao} \ Some studies of excitable networks have been inspired
by target and spiral waves in continuous media. \cite{SWSpiral}%
\cite{Showalter}\cite{Lewis}

In complex networks, loops are both generic and abundant.\cite{LoopsRRN}%
\cite{Loops} While short loops of length $L\ll N$ (where $N$ is the network
size) are rare in large random networks, ones with $L\gtrsim$ $\ln N$ occur
generically in numbers growing exponentially with $N$. \ Their number also
grows roughly exponentially with $L$ up to a maximum at a length $L_{m}\sim
N.$ \cite{LoopsRRN}\cite{Loops}\ 

In this letter we show that random excitable networks readily self-organize
into a CO state following a transient phase during which one or a few loops
are dynamically selected as driving (or pacemaker) loops. \ \ We describe the
mechanism of signal propagation and gain an understanding of the resulting
distributions of the oscillating states and associated driving loops. \ 

We consider networks of diffusively coupled, excitable elements with dynamics
described by the B\"{a}r model \cite{Baermodel}%
\begin{align}
\frac{du_{i}}{dt} &  =\frac{1}{\varepsilon}u_{i}(1-u_{i})(u_{i}-\frac{v_{i}%
+b}{a})+D%
{\displaystyle\sum\limits_{j=1}^{N}}
A_{ij}(u_{j}-u_{i}),\label{BaermodelEOM}\\
\frac{dv_{i}}{dt} &  =f(u_{i})-v_{i},\nonumber
\end{align}
where%
\[
f(u_{i})=\left\{
\begin{array}
[c]{cc}%
0 & u_{i}\leq1/3\\
1-6.75u_{i}(u_{i}-1)^{2} & 1/3<u_{i}<1\\
1 & u_{i}\geq1
\end{array}
\right.  ,
\]
$N$ is the number of nodes, $u_{i}$ and $v_{i}$ are dynamical variables, $D$
is the coupling strength, and $A_{ij}$ is the adjacency matrix. \ $a$, $b$,
and $\varepsilon$ are parameters, for which we adopt the values $a=0.84$,
$b=0.07$, and $\varepsilon=0.04$. \ In the absence of coupling, the individual
nodes display excitable dynamics, with a stable equilibrium at $(u,v)=(0,0)$
and an excitation threshold $u_{th}\approx0.1$. The dynamical equations were
integrated numerically using a fourth-order Runge-Kutta algorithm with time
step $\Delta t=0.1$. \ \ \ \ \ \ \ \ \ \ \ \ \ \ \ \ \ 

For the topology, we chose the undirected random regular network of degree $3
$ (RRN3), where nodes are randomly and symmetrically connected with the
constraint that all have the same degree $k=3$. \ The network size was $N=200
$. \ Other topologies and other values and distributions of $k$ will be
considered elsewhere. \ \ 

To excite the network, we used random initial conditions in which each node
was either displaced from equilibrium with probability $p$, or left at $(0,0)
$ with probability $1-p$. \ For the nodes that were displaced, \ initial
values of $u$ and $v$ were distributed randomly and independently within the
intervals $0.2\leq u\leq0.9$, $0\leq v\leq1$, thus placing them above the
excitation threshold but with some phase randomness. \ (This randomness proves
important, as discussed below.) \ After determining that the results were
largely insensitve to the value of $p$, we subsequently took $p=0.5$.\ 

\begin{figure}
[ptb]
\begin{center}
\includegraphics[
height=2.239in,
width=2.2131in
]%
{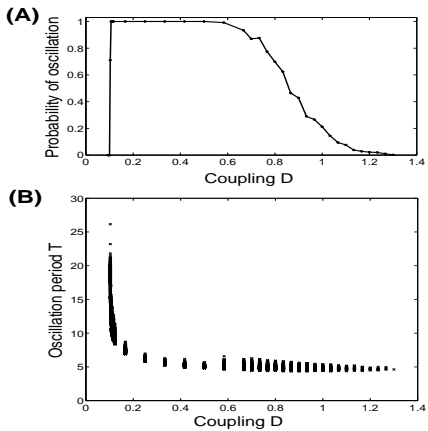}%
\caption{(A) Probability that a random initial condition with excitation
probability $p=0.5$ relaxes to an oscillatory state, averaged over 1000
initial conditions of 5 network realizations. \ (B) Scatter plot of
oscillation periods $T$. Periods are longer and more variable at weaker
coupling. \ }%
\label{couplingdependence}%
\end{center}
\end{figure}

\begin{figure}
[ptb]
\begin{center}
\includegraphics[
height=3.0623in,
width=2.5071in
]%
{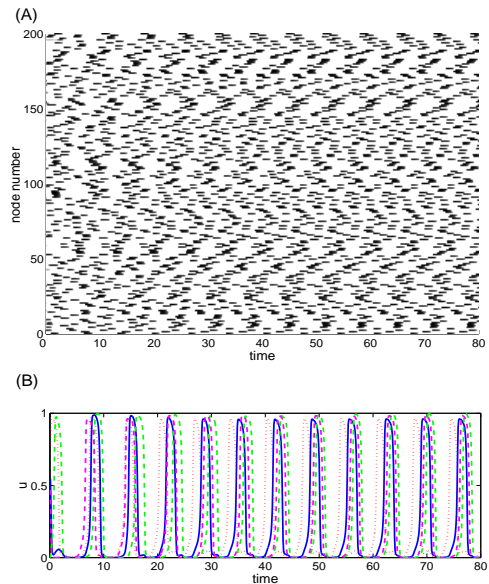}%
\caption{(color online): Example of a periodic network oscillation for
coupling strength $D=0.2$. \ (A) Raster plot of all nodes' activity. \ Gray
level is proportional to $u_{i}(t)$. \ The network quickly settles to a steady
oscillation in which all nodes fire with the same period in a fixed phase
relationship to each other. \ \ (B) \ $u(t)$ for an arbitrarily chosen node
(solid line) and its three neighbors (incoming: dotted and outgoing: dashed
lines). \ }%
\label{periodicexample}%
\end{center}
\end{figure}

Integrating the equations of motion (\ref{BaermodelEOM}), we found two
possible outcomes: \ either the system relaxed rapidly to the quiescent state
with all nodes at the fixed point or it reached a coherently oscillating state
(COS) like the one illustrated in fig.\ref{periodicexample}. \ In the COS, all
nodes fire at the same frequency, with a fixed phase relationship among them.
\ To examine the phase relations, we define a firing time $t_{i}$ of the $i$th
node as the time (interpolated linearly between discrete time steps) at which
$u_{i}$ crosses from below $0.5$ to above $0.5$. \ The interval between
successive firing times (interspike interval or ISI) converges to a stable
common value for all nodes, which is the oscillation period $T$. \ \ We
considered the system to have converged to a COS if the standard deviation of
the ISI's over the network remained below $10^{-4}$ for more than 100 time
units. As shown in fig.\ref{couplingdependence}, COS are highly probable for a
range of coupling strengths $0.1\lessapprox D\lessapprox0.6$, while the
periods decrease with increasing $D$.%

The local structure of the oscillatory state is illustrated by
fig.\ref{periodicexample}B, \ which shows the firing pattern of an arbitrarily
chosen node and its three neighbors. \ \ As each node fires once per
oscillation, one can define a time delay $\Delta_{ij}$ (with $-T/2\leq
\Delta_{ij}<T/2$) for each link in the network as the difference of firing
times $t_{i}-t_{j}$ between the nodes at its two ends, modulo the oscillation
period $T$. With the establishement of stable signed delays, the initially
undirected network self-organizes into a directed one in which the signal
propagates in only one direction along each link. \ Since each node $i$ must
be excited by one of its neighbors, it must have at least one incoming link
(positive $\Delta_{ij}$) but can have between $0$ and $2$ outgoing links
(negative $\Delta_{ij}$). \ The node illustrated in fig.\ref{periodicexample}B
is a \textquotedblleft diverging\textquotedblright\ node with one incoming and
two outgoing links. \ Since the total numbers of incoming and outgoing links
in the network must balance, all nodes cannot be converging. \ This implies
that CO can only occur if a single firing neighbor suffices to excite a node.
\ In this case, the firing of a node at the "upstream" end of a link
guarantees that the one at the "downstream" end will fire within a certain
time period, provided the downstream node is not refractory when it receives
the input. \ If the downstream node receives a second input before it fires,
\ it will be pulled over the threshold more quickly and fire sooner.
\ Converging inputs thus account for most of the variation in \ transmission
delays. \

\begin{figure}
[ptb]
\begin{center}
\includegraphics[
trim=0.000000in 0.000000in -0.107302in 0.074599in,
height=1.3266in,
width=1.567in
]%
{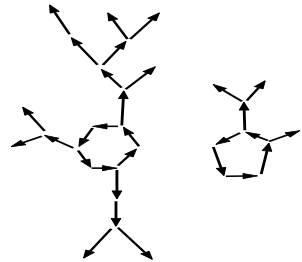}%
\caption{Branched circle structure of the pruned network. \ Following dominant
phase-advanced driving links backwards from any node inevitably leads to a
unidirectional loop from which the signal radiates along branches. \ \ }%
\label{bcdiag}%
\end{center}
\end{figure}

Liao et al. \cite{Liao} suggested a way of identifying the main transmission
pathways by selecting \ the so-called dominant phase-advanced driving (DPAD)
links, which then leads to identifying the pacemaker or driving loops
(DL).\ \ The DPAD for each node is the incoming link with the largest delay.
\ A justification for considering the earliest input to be the most important
is that one input is sufficient to guarantee the node's firing, and additional
inputs can only affect the timing \ \ Pruning the network to include only DPAD
links simplifies it to a so-called branched circle structure\cite{Liao}
consisting of trees attached to unidirectional DLs as illustrated in fig.
\ref{bcdiag}.

\begin{figure}
[ptb]
\begin{center}
\includegraphics[
height=3.9669in,
width=3.0424in
]%
{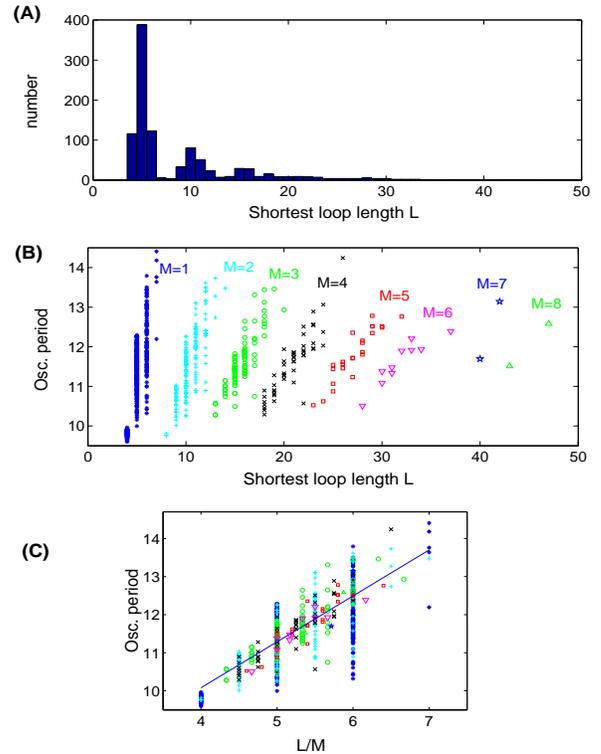}%
\caption{(color online): \ Pacemaker loops and oscillation periods for
$D=0.11$, for 1000 different initial conditions of 5 network realizations.
(Features are qualitatively similar for COS at $D=0.6$)\ (A) Histogram of
lengths of shortest driving loops. \ (The shape is qualitatively similar if
all loops are included.) \ A semilogarithmic plot of the cumulative
distribution (not shown here) reveals an exponentially decaying trend. \ (B)
Oscillation period $T$ vs. shortest loop length. \ Different symbols show
loops of different multiplicities M. \ Note that the clusters at different M
values correspond to the local peaks in panel (A). \ \ (C) The data from (B)
collapse into a single band when one plots $T$ vs. $L/M$. The best fit line
shown here is $T=1.21(L/M)+5.26$. \ For $D=0.6$, the corresponding best fit is
$T=0.25(L/M)+2.85$.}%
\label{perloop1}%
\end{center}
\end{figure}

To study the statistics of the COS's and their basins of attraction we
integrated eqs. (\ref{BaermodelEOM}) with $D=0.11$ for 1000 different initial
conditions (200 each for five different RRN3 realizations). \ All but one of
these initial conditions converged to a COS. \ We then measured the firing
delays and pruned the network as described above to identify the driving
loops. \ We found up to four distinct DLs in each COS, all mutually entrained
to oscillate with the same period. \ 50.9 percent of the COS had only one DL,
with the probability of more loops decreasing monotonically with number.\ \ In
each case, we measured the length $L$ of the shortest DL. \ The distribution
of these lengths is shown in figure \ref{perloop1}A. \ As figure
\ref{perloop1}B shows, \ the oscillation period $T$ is correlated with $L$,
but the data fall into clusters separated according to the number of pulses
circulating simultaneously around the loop, which we call the multiplicity
$M$. \ $M$ can be measured by adding up the transmission delays along the
driving loop to get the time for a single pulse to make one circuit, and
dividing this by the oscillation period. \ \ \ Fig.\ref{perloop1}C shows that
the data collapse into a single band when one plots $T$ vs. $L/M$. \ One can
interpret the slope of this band as the typical delay added by a link. \ A
notable feature of this plot is that the best fit line does not pass through
the origin as one would expect if waves of excitation travelled at a constant
velocity independent of the loop length. \ This feature becomes even more
noticeable when we examine data at stronger coupling (where transmission
delays are shorter). \ Data for $D=0.6$ look qualitatively similar to those
shown in figure \ref{perloop1}, \ except that the minimum value of $L$ is 6
rather than 4, and the slope of the plot of $T$ vs. $L/M$ is considerably
smaller, as expected, while the intercept is only slightly smaller. \ The
resulting overall spread of oscillation periods is correspondingly less. \ At
larger coupling strength, the transmission delay is smaller and evidently
plays less of a role in setting the period.%

Several features of our results merit discussion. \ First, we note the ease
with which the network is induced to oscillate: the probability of obtaining
oscillations from random initial conditions is nearly 1 within a range of
coupling strengths, and there is a large number of distinct oscillatory
attractors as attested by the scatter of periods and the variety of DLs.
\ Second, no external driving is needed to sustain CO activity, unlike the
case considered by \cite{Lewis}. \ Unlike cases where oscillation resulted
from adding directed shortcuts to a spatial network\cite{Showalter}%
\cite{SWSpiral}\cite{Roxin}, here the connections are symmetric. \ This
symmetry is only broken dynamically, producing different directed networks and
selecting different driving loops depending on the initial conditions. \ 

A non-trivial distribution of the initial conditions in phase space is
evidently crucial:\ initial conditions where one or more nodes were excited
synchronously (displaced to the same point above the excitation threshold)
failed to produce oscillations. \ \ A single excited node on a loop produces
two wavefronts of excitation propagating in opposite directions, which will
annihilate elsewhere on the loop. \ If an excited node adjoins a refractory
one, however, the signal is blocked from moving in one direction and a
unidirectional loop can be established.\cite{Lewis} \ Evidently the random
initial distribution we have used is sufficient to allow this. \ 

As shown in figure \ref{couplingdependence}, the probability of obtaining
oscillations from a random initial condition jumps rapidly from 0 to 1 at a
lower threshold coupling strength $D_{l}=0.1$, and falls off more gradually
for $D\gtrapprox0.6$. \ The lower threshold represents the minimum coupling
necessary for a node to be excited by the firing of a single neighbor. \ The
slower decay of the oscillation probability at $D\gtrapprox0.6$ can be
explained by the shortening of transmission delay compared to the refractory
period. \ In the cases where oscillation fails, it does so due to an avalanche
that spreads so rapidly through the network that almost all nodes are firing
at once. \ The activity burns out quickly when there are no nodes remaining
that are not already firing or refractory. \ 

The oscillation period is bounded from below by the refractory period. \ This
also implies a lower bound on the length of a DL depending on the maximum
delay per link: \ \ oscillation cannot be sustained if a pulse travelling
along a loop returns to a given node while that node is still refractory.
\ The observed lower bound on the loop size, $L_{\min}=4$ in the case
$D=0.11$, in fact increases to $6$ when $D$ is increased to $0.6$, because
increasing $D$ decreases the transmission delay. \ More generally, the above
argument explains the observed lower bound on the ratio $L/M$, which
represents the spacing between pulses on a DL. \ $L/M$, on the other hand, is
also apparently bounded from above, as evident from figure \ref{perloop1}C.
\ A possible explanation for this lies in the transient process by which a
particular loop becomes established as the primary DL. \ \ During this
transient phase, many pulses are propagating in an unsynchronized manner
through the network, and a number of potential pacemakers are competing. \ A
large gap between circulating pulses is therefore likely to be filled in.
\ This is analogous to oscillations in excitable spatial media, where
pacemakers with the shortest periods dominate.\cite{SWSpiral} \ 

The distribution of driving loops (figure \ref{perloop1}A) decays with $L$,
despite the fact that the number of loops present in the network \emph{grows}
exponentially with $L$. \ There is evidently a strong bias in favor of shorter
loops. This bias appears to be a statistical effect inherent in the topology
of the directed branched circle network rather than strictly a dynamical
effect. \ \ To check this hypothesis, we compared our results with a
\textquotedblleft null model\textquotedblright\ that generates branched circle
networks independently of the oscillatory dynamics. For the null model, we
generate directed networks as follows. \ First, assign a random direction to
each link in an RRN3, and then, where necessary, reverse directions of some
links to ensure that every node has at least one incoming link. \ For each
node, we then randomly choose one among its incoming links to be the dominant
one, and prune the others if the node is converging. \ The result is a network
satisfying the same topological constraints as the pruned network of DPAD
connections, but generated by a random process rather than a dynamical one.
\ The distribution of the shortest loop length $L$ for an ensemble of such
networks decays exponentially as it does in the DPAD network. \ Simply by
assigning directions and then pruning, one samples the loops of the original
undirected network unevenly. \ Longer loops have more chances either to fail
to be unidirectional when directions are assigned, or to be eliminated by
pruning. \ 

The driving loops uncovered by the DPAD reduction explain part but not all of
the variation in oscillation periods (see fig.\ref{perloop1}). \ Secondary
(non-dominant) connections alter the firing times, and are also responsible
for the entrainment of multiple DL when the latter occur. \ If the connections
that we prune for the sake of analysis were truly removed from the network, it
would be impossible for the resulting disconnected components to synchronize.
\ It is also worth emphasizing that when $M>1$, the multiple pulses are
precisely evenly spaced as they travel around the loop, so that the interval
between firings is constant for a given node. \ This is another form of
entrainment that we would not expect if the pruned connections were truly
absent. \ 

The dynamical organization of an undirected network giving rise to a directed
one with considerably different loop statistics provides an intriguing case
study of the distinction between "functional" networks defined by dynamical
interactions and the underlying structural networks of hard-wired connections.
\ Functional networks of observed correlations of activity among brain regions
have been studied as a tool to infer the underlying architecture\cite{BrainFN}%
, but the phenomenon discussed here illustrates that functional and structural
networks may have markedly different properties. \ 

\ Unlike the cases of interacting excitable cells where the oscillation is
treated as a collective phenomenon\cite{DeMonte-yeast}-\cite{Taylor2009}, in
the case of a fixed network specific signalling pathways can be more readily
identified. \  The networks studied here share some essential features with
brains, and it is plausible that the mechanism described here plays a role in
epileptic seizures.\cite{Epilepsy}\ 

\

\end{document}